\documentclass[12pt,a4paper]{article}
\usepackage{amsmath,amssymb,bm,ascmac,bbm,mathtools}
\usepackage[dvipdfmx, usenames]{color}
\usepackage[dvipdfmx]{graphicx}
\usepackage{xcolor}
\usepackage{here}
\usepackage{authblk}
\usepackage[normalem]{ulem}
\usepackage[hang,small,bf]{caption}
\usepackage[subrefformat=parens]{subcaption}
\usepackage{dcolumn}

\newcommand{\tr}{\text{tr}}

\newcommand{\mbb}{\mathbb}
\newcommand{\mfrak}{\mathfrak}

\newcommand{\ra}{\rangle}

\newcommand{\rank}{\text{rank}}

\newcommand{\hmu}{{\hat\mu}}

\newcommand{\mc}{\mathcal}
\newcommand{\uv}{\text{UV}}
\newcommand{\ir}{\text{IR}}
\newcommand{\wf}[1]{\widehat{\mfrak{#1}}}
\newcommand{\vecG}{\text{Vec}}
\newcommand{\vect}{\text{Vect}_{\mbb C}}
\newcommand{\svec}{\text{sVec}}
\newcommand{\fp}{\text{FPdim}}
\newcommand{\fib}{\text{Fib}}
\newcommand{\ising}{\text{Ising}}
\newcommand{\tc}{\text{ToricCode}}

\newcommand{\mods}{\text{mod }}
\newcommand{\rg}{\text{RG}}

\begin{document}
\title{Rational RG flow, extension, and Witt class}
\author{Ken KIKUCHI}
\affil{Department of Physics, National Taiwan University, Taipei 10617, Taiwan}
\date{}
\maketitle

\begin{abstract}
Consider a renormalization group flow preserving a pre-modular fusion category $\mathcal S_1$. If it flows to a rational conformal field theory, the surviving symmetry $\mathcal S_1$ flows to a pre-modular fusion category $\mathcal S_2$ with monoidal functor $F:\mathcal S_1\to\mathcal S_2$. By clarifying mathematical (especially category theoretical) meaning of renormalization group domain wall/interface or boundary condition, we find the hidden extended vertex operator (super)algebra gives a unique (up to braided equivalence) completely $(\mathcal S_1\boxtimes\mathcal S_2)'$-anisotropic representative of the Witt equivalence class $[\mathcal S_1\boxtimes\mathcal S_2]$. The mathematical conjecture is supported physically, and passes various tests in concrete examples including non/unitary minimal models, and Wess-Zumino-Witten models. In particular, the conjecture holds beyond diagonal cosets.

The picture also establishes the conjectured half-integer condition, which fixes infrared conformal dimensions mod $\frac12$. It further leads to the double braiding relation, namely braiding structures jump at conformal fixed points. As an application, we solve the flow from the $E$-type minimal model $(A_{10},E_6)\to M(4,3)$.
\end{abstract}

\makeatletter
\renewcommand{\theequation}
{\arabic{section}.\arabic{equation}}
\@addtoreset{equation}{section}
\makeatother

\section{Introduction and summary}
Problems in quantum field theory (QFT) are given in terms of renormalization group (RG) flows; problems are theories defined in ultraviolet (UV) and solutions are their behaviors in infrared (IR). For example, the unsolved quantum chromodynamics (QCD) also fits in this form. The theory is defined in UV as an $SU(3)$ gauge theory with matters (called quarks), and our goal is to identify its IR behaviors such as presence of gap, confinement, and spontaneous chiral symmetry breaking.

Usually, we have made the most of the symmetries preserved all along the RG flows. In particular, the 't Hooft anomaly matching \cite{tH79} has been extensively used to rule out trivially gapped IR behaviors.

In 2014, the definition of symmetry itself was generalized \cite{GKSW14}; symmetry operators are defined as topological operators. The definition reconciled some past observations \cite{V88,MS88,MS89,PZ00,FRS02,FRS03,FRS04a,FRS04b,FGS09,G12,BT17} into categorical symmetries. The 't Hooft anomalies in fusion categories are just part of the monoidal structures of fusion categories. The RG flows with fusion category symmetries are given by monoidal functors, and they ensure 't Hooft anomaly matching \cite{KKARG}. In addition, additional structures (such as braiding) in categorical symmetries succeeded to give mathematical (and physical) explanation, say, when and why which symmetries emerge in IR \cite{KK21,KKSUSY,KK22II,KK22free}. The categorical symmetries have tightened constraints on RG flows to give concrete conjectures \cite{KKSUSY,NK22,KKWZW,TN24}. 

Given these achievements, it is natural to wonder whether we could solve RG flows mathematically. In general, the attempt faces an immediate obstruction; we do not even know full symmetries preserved all along RG flows. There are, however, convenient classes of theories whose full symmetries are known, (diagonal\footnote{Most (if not all) non-diagonal RCFT can be obtained from diagonal ones by gauging or orbifolding. Furthermore, RG flows between non-diagonal RCFTs can be studied by diagonal ones using the commutativity of relevant operators and gauged symmetries. Therefore, in this paper, we limit ourselves to rational RG flows between diagonal RCFTs. In diagonal RCFTs, MFCs uniquely fix the theory. See, say, \cite{CKM17}.}) rational conformal field theories (RCFTs). Their symmetries are given by modular fusion categories (MFCs), and their relevant deformations preserve pre-modular fusion categories (pre-MFCs). (The definitions are collected in the Appendix \ref{preliminary}.) Manipulating this situation, with the hope in mind, an axiom of rational RG flows -- RG flows to RCFTs -- was proposed in \cite{KKARG}. However, the definition given in terms of Kan extension is too abstract to use it in practice to solve rational RG flows. It is desirable to have a description of (rational) RG flows which can actually be used to solve them. In this paper, we aim to give such a description. The picture is built on our main observation\\

\textbf{Conjecture.} \textit{Let $\mc S_\uv$ be the surviving pre-modular fusion category and $\mc S_\ir$ be its image pre-modular fusion subcategory in the infrared diagonal rational conformal field theory. The (simple) rational renormalization group flow stands in bijection with a unique (up to braided equivalence) completely $(\mc S_\uv\boxtimes\mc S_\ir)'$-anisotropic representative of the Witt equivalence class $[\mc S_\uv\boxtimes\mc S_\ir]$
\begin{equation}
    \{\text{Rational RG flow with }\mc S_\uv\}\cong\{\text{Completely }(\mc S_\uv\boxtimes\mc S_\ir)'\text{-anisotropic representative of }[\mc S_\uv\boxtimes\mc S_\ir]\}.\label{conjecture}
\end{equation}}\\

Note that conjecture claims a physical notion, rational RG flow, is in one-to-one correspondence with a mathematical notion, completely $(\mc S_\uv\boxtimes\mc S_\ir)'$-anisotropic representative (or connected étale algebra $A\in\mc S_\uv\boxtimes\mc S_\ir$ giving it) up to equivalence. Therefore, it enables us to solve rational RG flows by finding such mathematical objects. Luckily, the mathematical problem is manageable thanks to various techniques \cite{KO01,KL02,EP09,DMNO10,DNO11,EM21,CEM23,G23,KK23GSD,KK23preMFC,KK23rank5,KKH24,KK24rank9}. Some comments are in order.\\

\textbf{Remark.} If the surviving pre-MFC $\mc S_\uv$ is non-degenerate, i.e., modular, the conjecture claims the representative of $[\mc S_\uv\boxtimes\mc S_\ir]$ is completely anisotropic. Equivalently, the connected étale algebra $A\in\mc S_\uv\boxtimes\mc S_\ir$ giving it is the maximal one.\\

\textbf{Remark.} The unique (up to equivalence) existence of completely $(\mc S_\uv\boxtimes\mc S_\ir)'$-anisotropic representative of $[\mc S_\uv\boxtimes\mc S_\ir]$ signals the universality. Since universality is usually defined in terms of Kan extension, it would be interesting to see the explicit relation between the conjecture and the axiom \cite{KKARG}.\\

\textbf{Remark.} Connected étale algebras (with trivial topological twists) stand in bijection with extensions of vertex operator (super)algebras \cite{HKL14,CKM17}.\footnote{See, say, \cite{M21} for a concise review. We thank Yuto Moriwaki for bringing these references to our attention.} An extension of vertex operator algebra is a necessary condition to construct RG domain wall/interface or boundary condition à la Gaiotto \cite{G12}. In the paper, Gaiotto studied rational RG flows from $\text{RCFT}_\uv$ to $\text{RCFT}_\ir$. They have MFCs $\mc B_\uv,\mc B_\ir$, respectively. He found the product theory $\text{RCFT}_\uv\times\text{RCFT}_\ir$ has a hidden extended vertex operator (super)algebra. By the correspondence above, such extensions are in bijection with connected étale algebras with trivial topological twists in $\mc B_\uv\boxtimes\mc B_\ir$. The conjecture claims the `maximal' connected étale algebra leaving nontrivial transparent objects gives the rational RG flow.

Relatedly, when a rational RG flow admits RG domain wall/interface or boundary condition, the conjecture is trivially true by the bijection. Nontrivial statement of the conjecture is that \text{all} rational RG flows stand in bijection with completely $(\mc S_\uv\boxtimes\mc S_\ir)'$-anisotropic representative of $[\mc S_\uv\boxtimes\mc S_\ir]$.\\

\textbf{Remark.} Note that the conjecture is given \textit{not} in terms of the full UV symmetry, but in terms of only the surviving subsymmetry. Therefore, one does not have to know the full UV symmetry. Accordingly, the conjecture could be equally applied to rational RG flows from \textit{irrational} CFTs to RCFTs preserving pre-modular fusion subcategories. One interesting example is a relevant deformation of $c=1$ irrational CFT with a real coupling. Since all lower CFTs are rational \cite{KL02}, our conjecture should apply. It is important to test this class of rational RG flows to check our conjecture, but our tests are limited to rational UV theories.\\

While the observation should mathematically be viewed as a conjecture, it is supported by two physical arguments: (1) emergent symmetry and (2) minimization of free energy.

\begin{itemize}
    \item First, emergent symmetries are absent until an RG flow reaches the IR conformal fixed point. In addition, symmetries absent during an RG flow \textit{cannot} constrain the flow. Therefore, emergent symmetries should \textit{not} constrain RG flows. Here, full symmetries of (diagonal) RCFTs are given by MFCs. An MFC is in general a Deligne tensor products of prime MFCs \cite{DMNO10}. On the other hand, relevant deformations of UV RCFTs preserve possibly degenerate pre-MFCs. Therefore, centers of degenerate pre-MFCs enhance to prime MFCs in rational RG flows \cite{KK21,KKSUSY,KK22II}. Now, suppose a connected étale algebra corresponding to a rational RG flow contains nontrivial transparent simple objects. Then, the auxiliary theory in Gaiotto's construction (we call the theory an RG theory for simplicity) affects emergent symmetries, a contradiction. We learn connected étale algebras (or equivalently extensions of vertex operator (super)algebras) should be blind to nontrivial transparent objects in the surviving pre-MFCs.
    \item Second, the largest Frobenius-Perron dimension $\fp_{\mc S_\uv\boxtimes\mc S_\ir}(A)$ of a connected étale algebra $A$ is energetically favored. To see this, we first recall the largest Frobenius-Perron dimension minimizes the Frobenius-Perron dimension of the category $(\mc S_\uv\boxtimes\mc S_\ir)_A^0$ of dyslectic (right) $A$-modules. More precisely, in order to use the formula (\ref{FPdimformula}), we first remove the transparent simple objects. The resulting pre-MFC $\mc B$ no longer have nontrivial transparent simple objects (mathematically, this is denoted $A\cap(\mc S_\uv\boxtimes\mc S_\ir)'=1$), and it is modular. Since the connected étale algebra does not contain nontrivial transparent simple object, $A$ remains intact. For the MFC $\mc B$, we can apply the formula to get the smallest Frobenius-Perron dimension $\fp(\mc B)$. In pseudo-unitary fusion categories, the Frobenius-Perron dimension equals categorical dimension $\fp(\mc B)=D(\mc B)^2$. An MFC with categorical dimension $D$ is known to contribute entropy \cite{KP05,LW05} by $-\ln D$. At temperature $T$ (defined as the length $1/T$ of the compactified Euclidean time), an MFC with categorical dimension $D$ thus contributes free energy by
    \[ F\ni T\ln D. \]
    Physical processes are realized to minimize free energies. Therefore, if there are various possible MFCs with different categorical dimensions, the MFC with the smallest categorical dimension is physically favored. (This physical principle succeeded to explain or even predict which MFCs emerge in IR \cite{KK22free}.) Therefore, minimization of free energies favors the largest Frobenius-Perron dimension $\fp_{\mc S_\uv\boxtimes\mc S_\ir}(A)$.
\end{itemize}

The picture based on extension and vertex operator (super)algebra also proves the conjectured half-integer condition \cite{KKWZW}:\\

\textbf{Theorem.} \textit{Let $(\mc S_\uv,c^\uv)$ be the surviving pre-modular fusion subcategory and $(\mc S_\ir,c^\ir)$ be its image pre-modular fusion subcategory in the infrared diagonal rational conformal field theory under monoidal functor $F:\mc S_\uv\to\mc S_\ir$. If the rational renormalization group flow is described by renormalization group domain wall/interface or boundary condition, then surviving symmetry objects $s_j^\uv\in\mc S_\uv, s_j^\ir=F(s_j^\uv)$ obey half-integer condition
\begin{equation}
    h_j^\uv+h_j^\ir\in\frac12\mbb Z,\label{halfintegercond}
\end{equation}
and hence the double braiding relation
\begin{equation}
    \forall s_i^\uv,s_j^\uv\in\mc S_\uv,\quad c^\ir_{s^\ir_j,s^\ir_i}\circ c^\ir_{s^\ir_i,s^\ir_j}=(c^\uv_{s^\uv_j,s^\uv_i}\circ c^\uv_{s^\uv_i,s^\uv_j})^*,\label{doublebraidingrelation}
\end{equation}
where the right-hand side is a complex conjugate.}\\

\textit{Proof.} When a rational RG flow is described by RG domain wall/interface or boundary condition, we have product symmetry objects $s^\uv_js^\ir_j\in\mc S:=\mc S_\uv\boxtimes\mc S_\ir$ of a surviving symmetry object $s^\uv_j$ and its image $s^\ir_j=F(s^\uv_j)$ in the product theory. Because of the fusion product $s^\uv_js^\ir_j\otimes1\cong s^\uv_js^\ir_j$, there exist chiral disorder operators at the end of the product symmetry object. Let $(h_i,h_j)$ be a conformal dimension with holomorphic $h_i$ and anti-holomorphic conformal dimension $h_j$. Then, the disorder operators have conformal dimensions $(h^\uv_j+h^\ir_j,0),(0,h^\uv_j+h^\ir_j)$. For the product symmetry object to be able to end topologically on the RG boundary condition in a unique way \cite{G12}, the chiral disorder operators should have single-valued correlation functions.\footnote{We thank Ingo Runkel and Yifan Wang for pointing out errorneous argument in the previous version.} This requires
\begin{equation}
    h^\uv_j+h^\ir_j\in\frac12\mbb Z,\label{singlevalued}
\end{equation}
proving the half-integer condition (\ref{halfintegercond}).

Furthermore, Gaiotto argued that the chiral disorder operators define extended vertex operator (super)algebras \cite{G12}.\footnote{In this viewpoint, weights of vertex operator superalgebras take values in $\frac12\mbb Z$. This gives more axiomatic proof of the half-integer condition. We thank Yuto Moriwaki for discussion on this point.} Vertex operator superalgebras have natural $\mbb Z/2\mbb Z$-grading assigning zero to even part $\mc S^{\bar0}$ and one to odd part $\mc S^{\bar1}$. The even part has integer weights, and odd part has half-integer weights. These satisfy the assumptions of the theorem in \cite{KKWZW}, and hence the double braiding relation follows. $\square$\\

\textbf{Remark.} Not all rational RG flows may admit RG domain wall/interface or boundary condition. Although we do not know any, a `counterexample' does not have to obey the half-integer condition nor double braiding relation if it existed. However, Gaiotto claimed his construction holds more generally. At least, physical picture of conformal interface and conformal boundary condition seems to hold quite generally. Therefore, such `counterexamples' seem unlikely to exist.\\

\textbf{Remark.} Physicists should be surprised by the double braiding relation. While braiding structures are proved to be discrete \cite{ENO02} (known as Ocneanu rigidity), the relation says the UV braiding structure jumps to a \textit{different} IR braiding structure under a rational RG flow. This should be viewed as a counterexample to physicists' argument ``discrete quantities cannot change under continuous transformations.''\\

\textbf{Remark.} Relatedly, monoidal functors giving rational RG flows are \textit{not} braided. This property could depend on spacetime dimensions. For example, \cite{BBR24} claimed monoidal functors in three dimensions are braided.\\

\textbf{Remark.} The theorem fixes IR conformal dimensions (mod $\frac12$), part of conformal data. To the best of our knowledge, this would be the first analytic (and exact mod $\frac12$) constraint on conformal data. The conformal dimensions strongly constrain central charges of modular fusion subcategories in $\mc S_\ir$. The constraint together with additional constraints such as the $c$-theorem \cite{cthm} (or $c^\text{eff}$-theorem \cite{CDR17} for non-unitary cases) or minimization of free energy \cite{KK22free} could fix an IR MFC uniquely. That would realize our hope mentioned above. We leave detailed study in this direction to future, but discuss a simple application on the RG flow from the $E$-type minimal model $(A_{10},E_6)$ in the next section.\\

\textbf{Remark.} Some applications of the double braiding relation were discussed in \cite{KK22II}. In particular, it could analytically rule out massless IR phases. An example $M(6,5)+\epsilon'$ was discussed in the paper.\\

In the next section, we check our conjecture in concrete examples. We will see the half-integer conditions are beautifully satisfied, as they should be. We would also like to mention that in all examples, conformal dimensions decrease ``monotonically,'' $h^\ir_j<h^\uv_j$. This is natural from the Wilsonian viewpoint \cite{KK22II}. The examples we study are summarized in the following\\
\begin{table}[H]
\begin{center}
\makebox[1 \textwidth][c]{       
\resizebox{1.2 \textwidth}{!}{\begin{tabular}{c|c|c|c|c|c}
    Rational RG flow&Preserved $\mc S_\uv$&$\mc S:=\mc S_\uv\boxtimes\mc S_\ir$&$\mc S'$&Representative of $[\mc S]$&Section\\\hline\hline
    $M(5,4)\to M(4,3)$&$\ising$&$\ising\boxtimes\ising$&$\vect$&$\tc$&\ref{M54toM43}\\\hline
    $M(6,5)\to M(5,4)$&$\fib\boxtimes\vecG_{\mbb Z/2\mbb Z}^1$&$\fib\boxtimes\vecG_{\mbb Z/2\mbb Z}^1\boxtimes\fib\boxtimes\svec$&$\vecG_{\mbb Z/2\mbb Z}^1\boxtimes\svec$&$\vecG_{\mbb Z/2\mbb Z}^1\boxtimes\svec$&\ref{M65toM54}\\\hline
    $M(7,6)\to M(6,5)$&$su(2)_4$&$su(2)_4\boxtimes su(2)_4$&$\vect$&$\vect$&\ref{M76toM65}\\\hline
    $M(8,7)\to M(7,6)$&$\svec\boxtimes psu(2)_5$&$\svec\boxtimes psu(2)_5\boxtimes\vecG_{\mbb Z/2\mbb Z}^1\boxtimes psu(2)_5$&$\svec\boxtimes\vecG_{\mbb Z/2\mbb Z}^1$&$\svec\boxtimes\vecG_{\mbb Z/2\mbb Z}^1$&\ref{M87toM76}\\\hline
    $M(3,5)\to M(2,5)$&$\fib$&$\fib\boxtimes\fib$&$\vect$&$\vect$&\ref{M35toM25}\\\hline
    $M(15,4)\to M(9,4)$&$\ising$&$\ising\boxtimes\ising$&$\vect$&$\tc$&\ref{M154toM94}\\\hline
    $\wf{su}(3)_2\to\wf{su}(3)_1$&$\vecG_{\mbb Z/3\mbb Z}^1$&$\vecG_{\mbb Z/3\mbb Z}^1\boxtimes\vecG_{\mbb Z/3\mbb Z}^1$&$\vect$&$\vect$&\ref{su32tosu31}\\\hline
    $(\wf{e_7})_3\to(\wf{e_7})_1$&$\vecG_{\mbb Z/2\mbb Z}^{-1}$&$\vecG_{\mbb Z/2\mbb Z}^{-1}\boxtimes\vecG_{\mbb Z/2\mbb Z}^{-1}$&$\vect$&$\vect$&\ref{e73toe71}\\\hline
    $(A_{10},E_6)\to M(4,3)$&$\ising$&$\ising\boxtimes\ising$&$\vect$&$\vect$&\ref{A10E6toM43}
\end{tabular}.}}
\end{center}
\caption{Summary of examples}
\end{table}

\section{Tests}
In this section, we expose our conjecture to tests. We start from the class of rational RG flows where RG domain walls/interfaces surely exist. Namely, we first study rational RG flows between unitary discrete series of minimal models $M(m+1,m)\to M(m,m-1)$. Building up on the success, we next study known rational RG flows between non-unitary minimal models, and Wess-Zumino-Witten models. Finally, we apply our machinery to RG flows from the $E$-type minimal model $(A_{10},E_6)$.

Throughout the section, we use the following notation. Simple objects in the surviving UV symmetry category $\mc S_\uv$ are denoted with lower cases such as $x,y,z,\dots$, and simple objects in image IR symmetry category $\mc S_\ir$ are denoted with upper cases such as $X,Y,Z,\dots$. We use the same letter for objects to be identified. For example, $X=F(x),Y=F(y),\dots$ where $F:\mc S_\uv\to\mc S_\ir$ is the monoidal functor giving the rational RG flow.

\subsection{Unitary discrete series}
For small $m$, both symmetry categories and connected étale algebras in them have been studied extensively. Thus, it turns out that almost all we should check is the categorical dimension of $(\mc S_\uv\boxtimes\mc S_\ir)_A^0$. The dimension can be computed as follows. The category $(\mc S_\uv\boxtimes\mc S_\ir)_A^0$ is a pre-modular fusion subcateogry of the MFC $\mc B_\rg$ describing an RG theory. (Remember that we call the auxiliary theory à la Gaiotto with extended vertex operator (super)algebra an RG theory.) In cases of diagonal cosets, the RG theories were also identified. For the RG flow $M(m+1,m)=\frac{\wf{su}(2)_{m-2}\times\wf{su}(2)_1}{\wf{su}(2)_{m-1}}\to M(m,m-1)=\frac{\wf{su}(2)_{m-3}\times\wf{su}(2)_1}{\wf{su}(2)_{m-2}}$, the RG theory is the coset
\[ \frac{\wf{su}(2)_{m-3}\times\wf{su}(2)_1\times\wf{su}(2)_1}{\wf{su}(2)_{m-1}}. \]
The theory has the modular $S$-matrix
\[ S_{[\hmu,\hat\nu,\hat\rho;\hat\sigma],[\hmu',\hat\nu',\hat\rho;\hat\sigma']}=2\frac{(-1)^{\nu_1\nu_1'+\rho_1\rho_1'}}{\sqrt{(m-3)(m-1)}}\sin\frac{\pi(\mu_1+1)(\mu_1'+1)}{m-3}\sin\frac{\pi(\sigma_1+1)(\sigma_1'+1)}{m-1}. \]
Therefore, the RG theory has the categorical dimension
\begin{equation}
    D(\mc B_\rg)=\frac1{S_{11}}=\frac{\sqrt{(m-1)(m+1)}}{2\sin\frac\pi{m-1}\sin\frac\pi{m+1}}.\label{DBRG}
\end{equation}
The categorical dimension includes contributions from broken UV symmetry objects and emergent IR symmetry objects. In order to remove the contributions, we first divide by categorical dimensions of these MFCs, and next multiply by categorical dimensions coming from transparent simple objects in $\mc S:=\mc S_\uv\boxtimes\mc S_\ir$.

Since the class of RG flows have periodic patter with four in $m$, it is enough to study four $m$'s. We start from small $m$.

\subsubsection{$M(5,4)+\phi_{1,3}\to M(4,3)$.}\label{M54toM43}
The UV symmetry category is given by
\[ \mc B_\uv\simeq\fib\boxtimes\ising=\{1,z\}\boxtimes\{1,x,y\}. \]
The relevant deformation with $\phi_{1,3}$ preserves modular fusion subcategory
\[ \mc S_\uv\simeq\ising. \]
It flows to the IR, and is mapped to
\[ \mc B_\ir(=\mc S_\ir)\simeq\ising=\{1,X,Y\}. \]
From their fusion products, we can uniquely identify symmetry objects
\begin{equation}
\begin{array}{ccccc}
M(5,4):&1_0&x_{\frac32}&y_{\frac7{16}}\\
&\downarrow&\downarrow&\downarrow\\
M(4,3):&1_0&X_{\frac12}&Y_{\frac1{16}}
\end{array}.\label{m4}
\end{equation}
For convenience, we added conformal dimensions in subscripts. We see the half-integer condition (\ref{halfintegercond}) is satisfied:
\[ (h_1,h_{xX},h_{yY})=(0,2,\frac12). \]
In our $m=4$, the categorical dimension (\ref{DBRG}) is given by
\[ D(\mc B_\rg)=\sqrt{10+2\sqrt5}. \]
Removing the contributions from broken $\fib$, we get $D(\mc S_A^0)=2$, or
\[ \fp(\mc S_A^0)=4. \]
Our conjecture claims the category is the category of dyslectic (right) $A$-modules for a connected étale algebra $A\in\mc S\simeq\ising\boxtimes\ising$. (With this identification in mind, we already wrote the category $\mc S_A^0$.) Such connected étale algebras have been classified in \cite{KK24rank9}. For the specific conformal dimensions (and quantum dimensions), the maximal connected étale algebra is
\[ A\cong 1\oplus xX. \]
This gives
\[ \mc S_A^0\simeq\tc, \]
matching the Frobenius-Perron dimension. The resulting Toric Code MFC $\mc S_A^0$ has four simple objects with conformal dimensions $0,\frac12,\frac12,\frac12$ mod 1. It is completely anisotropic \cite{KK23rank5}. Since $\mc S\simeq\ising\boxtimes\ising$ is non-degenerate, this example supports our conjecture.

\subsubsection{$M(6,5)+\phi_{1,3}\to M(5,4)$.}\label{M65toM54}
The UV RCFT has symmetry category
\[ \mc B_\uv\simeq\fib\boxtimes su(2)_4=\{1,x\}\boxtimes\{1,y,z,s,t\}. \]
The relevant operator $\phi_{1,3}$ preserves pre-modular fusion subcategory
\[ \mc S_\uv\simeq\fib\boxtimes\vecG_{\mbb Z/2\mbb Z}^1 \]
with $\mc S_\uv'\simeq\vecG_{\mbb Z/2\mbb Z}$. It flows to
\[ \mc S_\ir\simeq\fib\boxtimes\svec, \]
and enhances to
\[ \mc B_\ir\simeq\fib\boxtimes\ising=\{1,X\}\boxtimes\{1,Y,W\} \]
in order to make the IR symmetry category modular \cite{KK21}. The simple objects flow as
\begin{equation}
\begin{array}{cccccc}
M(6,5):&1_0&x_{\frac75}&y_3&xy_{\frac25}\\
&\downarrow&\downarrow&\downarrow&\downarrow\\
M(5,4):&1_0&X_{\frac35}&Y_{\frac32}&XY_{\frac1{10}}
\end{array}.\label{m5}
\end{equation}
Again, half-integer conditions are satisfied
\[ (h_1,h_{xX},h_{yY},h_{xyXY})=(0,2,\frac92,\frac12). \]
For $m=5$, the categorical dimension is given by
\[ D(\mc B_\rg)=4\sqrt3. \]
Removing contributions from broken $su(2)_4$ and emergent $\ising$, and multiplying contributions from transparent simple objects, we get
\[ D(\mc S_A^0)=\frac{4\sqrt3}{\sqrt{12}\sqrt4}\times\sqrt{2^2}=2, \]
or
\[ \fp(\mc S_A^0)=4. \]
(Recall $\mc S:=\mc S_\uv\boxtimes\mc S_\ir$.) Since our connected étale algebra obeys $A\cap\mc S'=1$, we need to find connected étale algebras in $\fib\boxtimes\fib$. The classification was done in \cite{BD11,KK23rank5}, and the maximal connected étale algebra is given by
\[ A\cong 1\oplus xX. \]
Actually, it is a Lagrangian algebra, and $(\fib\boxtimes\fib)_A^0\simeq\vect$. Putting back the transparent simple objects, we find
\[ \mc S_A^0\simeq\vecG_{\mbb Z/2\mbb Z}^1\boxtimes\svec\simeq\mc S'. \]
This matches the Frobenius-Perron dimension. Furthermore, it is completely $\mc S'$-anisotropic. The example also supports our conjecture.

\subsubsection{$M(7,6)+\phi_{1,3}\to M(6,5)$.}\label{M76toM65}
The UV MFC is given by
\[ \mc B_\uv\simeq psu(2)_5\boxtimes su(2)_4=\{1,s,t\}\boxtimes\{1,x,y,z,w\}. \]
The relevant deformation with $\phi_{1,3}$ preserves modular fusion subcategory
\[ \mc S_\uv\simeq su(2)_4. \]
It flows to the IR subsymmetry
\[ \mc S_\ir\simeq su(2)_4, \]
and enhances to
\[ \mc B_\ir\simeq\fib\boxtimes su(2)_4=\{1,U\}\boxtimes\{1,X,Y,Z,W\} \]
to make (additive) central charge consistent with the $c$-theorem \cite{KK21}. The simple objects are identified as
\begin{equation}
\begin{array}{cccccc}
M(7,6):&1_0&x_0&y_{\frac38}&z_{\frac{23}8}&w_{\frac43}\\
&\downarrow&\downarrow&\downarrow&\downarrow&\downarrow\\
M(6,5):&1_0&X_0&Y_{\frac18}&Z_{\frac{13}8}&W_{\frac23}
\end{array}.\label{m6}
\end{equation}
We see the half-integer conditions are beautifully satisfied:
\[ (h_1,h_{xX},h_{yY},h_{zZ},h_{wW})=(0,0,\frac12,\frac92,2). \]
For $m=6$, the categorical dimension is
\[ D(\mc B_\rg)=\sqrt{\frac{7(5+\sqrt5)}{2\sin^2\frac\pi7}}. \]
Removing the contributions from broken $psu(2)_5$ and emergent $\fib$, we get
\[ D(\mc S_A^0)=\sqrt{\frac{7(5+\sqrt5)}{2\sin^2\frac\pi7}}/(\sqrt{\frac7{4\sin^2\frac\pi7}}\sqrt{\frac{5+\sqrt5}2})=2, \]
or
\[ \fp(\mc S_A^0)=4. \]
With inspection, we find a candidate
\[ A\cong1\oplus xX\oplus wW. \]
Indeed, in view of anyon condensation \cite{BSS02,BSS02',BS08}, one finds $A\in\mc S$ can be condensed.\footnote{The simple objects of $\mc S_A$ are given by
\begin{align*}
    m_1\cong A,\quad m_2\cong x\oplus X\oplus wW,\quad&m_3\cong yY\oplus zZ,\quad m_4\cong yZ\oplus zY,\\
    m_5\cong y\oplus zX\oplus yW\oplus zW,\quad m_6\cong Y\oplus xZ\oplus wY\oplus wZ,\quad&m_7\cong z\oplus yX\oplus yW\oplus zW,\quad m_8\cong Z\oplus xY\oplus wY\oplus wZ,\\
    m_9\cong w\oplus W\oplus wX\oplus xW\oplus wW,\quad&m_{10}\cong yY\oplus zZ\oplus yZ\oplus zY.
\end{align*}
The first four simple objects are deconfined, and the other six are confined. The deconfined simple objects have conformal dimensions $0,0,\frac12,0$ mod 1. We also found a non-negative integer matrix representation (NIM-rep), but we omit them.} The result tells us
\[ \mc S_A^0\simeq\tc, \]
matching the Frobenius-Perron dimension. This implies $A$ is connected étale \cite{K13}. Is the category $\mc S_A^0$ completely anisotropic? No. Connected étale algebras in Toric Code MFC have been classified in \cite{KK23rank5}. The category has two nontrivial connected étale algebras
\[ A'\cong m_1\oplus m_2,\quad m_1\oplus m_4. \]
They both lead to the category\footnote{This further extension caused by a connected étale algebra is consistent with the blindness of them to nontrivial transparent simple objects because in this example $\mc S'\simeq\vect$. Equivalently, one could realize the extension in one step by choosing a Lagrangian algebra.} of dyslectic (right) $A'$-modules
\[ (\mc S_A^0)_{A'}^0\simeq\vect. \]
The trivial $\vect$ is completely anisotropic. By (\ref{BBA0Wittequivalent}), we have
\[ [(\mc S_A^0)_{A'}^0]=[\mc S_A^0]=[\mc S]. \]
Therefore, $[(\mc S_A^0)_{A'}^0]$ gives a unique (up to braided equivalence) completely anisotropic representative of the Witt equivalence class $[\mc S]$. This example further supports our conjecture.\footnote{In order to make this example mathematically rigorous, we still need to prove $A\cong 1\oplus xX\oplus wW$ is a (connected) étale algebra. Since $F$-symbols of $su(2)_4$ seem unknown, we could not prove this. However, mathematicians would be able to prove this possibly with some codes.}

\subsubsection{$M(8,7)+\phi_{1,3}\to M(7,6)$.}\label{M87toM76}
The UV MFC is
\[ \mc B_\uv\simeq psu(2)_5\boxtimes su(2)_6=\{1,x,y\}\boxtimes\{1,z,s,t,u,v,w\}. \]
The relevant operator $\phi_{1,3}$ commutes with pre-modular fusion subcategory
\[ \mc S_\uv\simeq\svec\boxtimes psu(2)_5. \]
It flows to the IR subsymmetry
\[ \mc S_\ir\simeq\vecG_{\mbb Z/2\mbb Z}^1\boxtimes psu(2)_5, \]
and enhances to
\[ \mc B_\ir\simeq psu(2)_5\boxtimes su(2)_4=\{1,X,Y\}\boxtimes\{1,Z,P,Q,R\} \]
to make it modular \cite{KK21}. The simple objects are identified as
\begin{equation}
\begin{array}{ccccccc}
M(8,7):&1_0&x_{\frac{34}7}&y_{\frac97}&z_{\frac{15}2}&xz_{\frac5{14}}&yz_{\frac{39}{14}}\\
&\downarrow&\downarrow&\downarrow&\downarrow&\downarrow&\downarrow\\
M(7,6):&1_0&X_{\frac{22}7}&Y_{\frac57}&Z_5&XZ_{\frac17}&YZ_{\frac{12}7}
\end{array}.\label{m7}
\end{equation}
We again see the half-integer condition is perfectly satisfied:
\[ (h_1,h_{xX},h_{yY},h_{zZ},h_{xzXZ},h_{yzYZ})=(0,8,2,\frac{25}2,\frac12,\frac92). \]

Let us identify $\mc S_A^0$. For $m=7$, the categorical dimension (\ref{DBRG}) is
\[ D(\mc B_\rg)=4\sqrt{6(2+\sqrt2)}. \]
Removing the contributions from the broken $su(2)_6$ and emergent $su(2)_4$, and multiplying by those from transparent simple objects, we get
\[ D(\mc S_A^0)=4\sqrt{6(2+\sqrt2)}/(\sqrt{8(2+\sqrt2)}\sqrt{12})\times\sqrt{2^2}=2, \]
or
\[ \fp(\mc S_A^0)=4. \]
Since the connected étale algebra obeys $A\cap\mc S'=1$, we only have to look at $psu(2)_5\boxtimes psu(2)_5$. Connected étale algebras in the rank nine MFC have been classified in \cite{KK24rank9}. For our MFC, since the IR MFC has the reverse braiding of the UV MFC, the maximal connected étale algebra is the Lagrangian algebra
\[ A\cong1\oplus xX\oplus yY. \]
Putting back the transparent simple objects, we find
\[ \mc S_A^0\simeq\svec\boxtimes\vecG_{\mbb Z/2\mbb Z}^1\simeq\mc S'. \]
This identification matches the Frobenius-Perron dimension. Furthermore, it is completely $\mc S'$-anisotropic. This example gives further support of our conjecture.

\subsection{Non-unitary minimal models}
The examples above belong to the class studied by Gaiotto. In particular, it is known that they admit RG domain wall/interface or boundary condition. However, he claimed the description holds more generally. We find our conjecture holds in wider class of rational RG flows. In this subsection, we test our conjecture in rational RG flows between non-unitary minimal models \cite{Z90,Z91,M91,RST94,DDT00,TN24}.

\subsubsection{$M(3,5)+\phi_{1,2}\to M(2,5)$.}\label{M35toM25}
The UV MFC is
\[ \mc B_\uv\simeq\vecG_{\mbb Z/2\mbb Z}^{-1}\boxtimes\fib=\{1,y\}\boxtimes\{1,x\}. \]
The relevant operator $\phi_{1,2}$ preserves modular fusion subcategory
\[ \mc S_\ir\simeq\fib. \]
It flows to IR subsymmetry
\[ \mc S_\ir\simeq\fib. \]
Since the category is consistent, no emergent symmetry objects are needed \cite{KK22free}. Indeed, it is the full IR symmetry
\[ \mc B_\ir\simeq\fib=\{1,X\}. \]
The symmetry objects are identified as
\begin{equation}
\begin{array}{ccc}
M(3,5):&1_0&x_{\frac15}\\
&\downarrow&\downarrow\\
M(2,5):&1_0&X_{-\frac15}
\end{array}.\label{35to25}
\end{equation}
We see the half-integer condition is satisfied:
\[ (h_1,h_{xX})=(0,0). \]

Let us identify $\mc S_A^0$. In this case, we do not have a formula for $D(\mc S_A^0)$. However, connected étale algebras in $\mc S:=\mc S_\uv\boxtimes\mc S_\ir\simeq\fib\boxtimes\fib$ have been classified in \cite{BD11,KK23rank5}. For an RG theory to enjoy extended vertex operator (super)algebra, $A$ should be nontrivial. Luckily, there is only one nontrivial connected étale algebra
\[ A\cong1\oplus xX. \]
Actually, it is a Lagrangian algebra, and we find
\[ \mc S_A^0\simeq\vect. \]
It is completely anisotropic. This non-unitary example also supports our conjecture beyond those studied in \cite{G12}.

\subsubsection{$M(15,4)+\phi_{1,7}\to M(9,4)$.}\label{M154toM94}
Let us also study a newly suggested rational RG flow \cite{TN24}, $M(15,4)\to M(9,4)$. The UV theory has MFC
\[ \mc B_\uv\simeq\ising\boxtimes psu(2)_{13}=\{1,x,y\}\boxtimes\{1,z,s,t,u,v,w\}. \]
The relevant operator $\phi_{1,7}$ preserves modular fusion subcategory
\[ \mc S_\uv\simeq\ising. \]
In \cite{TN24}, based on spin constraint \cite{KKSUSY}, they conjectured the IR theory can be the non-unitary minimal model $M(9,4)$ (for a certain relevant coupling). In the scenario, the surviving symmetries flow to
\[ \mc S_\ir\simeq\ising=\{1,X,Y\}. \]
The simple objects are identified as
\begin{equation}
\begin{array}{cccc}
M(15,4):&1_0&x_{\frac{13}2}&y_{\frac{37}{16}}\\
&\downarrow&\downarrow&\downarrow\\
M(9,4):&1_0&X_{\frac72}&Y_{\frac{19}{16}}
\end{array}.\label{154to94}
\end{equation}
The surviving simple objects obey the half-integer condition:
\[ (h_1,h_{xX},h_{yY})=(0,10,\frac72). \]
Again, we do not know the categorical dimension of $\mc S_A^0$, however, the connected étale algebras in $\mc S:=\mc S_\uv\boxtimes\mc S_\ir\simeq\ising\boxtimes\ising$ have been classified in \cite{KK24rank9}. In our MFC $\mc S$, there are two connected étale algebras $A\cong1,1\oplus xX$. For an RG theory to have extended vertex operator (super)algebra, the rational RG flow should correspond to the nontrivial one
\[ A\cong1\oplus xX. \]
The resulting category $\mc S_A^0$ of dyslectic (right) $A$-modules is
\[ \mc S_A^0\simeq\tc \]
with conformal dimensions $0,\frac12,\frac12,\frac12$ mod 1. It is completely anisotropic \cite{KK23rank5}. This new rational RG flow gives further support of our conjecture.

\subsection{Wess-Zumino-Witten model}
This is another class of RCFTs. Since we need some backgrounds later, let us briefly review necessary facts.

Let $\mfrak g$ be a simple Lie algebra with rank $r$ and choose a level $k\in\mbb N$. The Wess-Zumino-Witten (WZW) model $\wf g_k$ has central charge
\begin{equation}
    c=\frac{k\dim\mfrak g}{k+g},\label{cWZW}
\end{equation}
where $g$ is the dual Coxeter number. The primary operators (and hence symmetry objects) are labeled by affine weights $\hmu$ in the affine dominant weights
\begin{equation}
    P^k_+:=\left\{\hmu|0\le\mu_j\&0\le\sum_{j=1}^ra_j^\vee\mu_j\le k\right\}.\label{Pk+}
\end{equation}
Here, $a_j^\vee$'s are comarks (a.k.a. dual Kac labels). A primary operator $\phi_\hmu$ has conformal dimension
\begin{equation}
    h_\hmu=\frac{(\mu,\mu+2\rho)}{2(k+g)},\label{hmu}
\end{equation}
where $\rho=(1,\dots,1)$ is the Weyl vector (a.k.a. principal vector). The scalar product in the numerator is given by
\[ (\mu,\lambda)=\sum_{i,j=1}^r\mu_i\lambda_jF_{ij}, \]
where the quadratic form matrix
\[ F_{ij}:=(\omega_i,\omega_j)=F_{ji} \]
can be found in the Appendix 13.A of the yellow book \cite{FMS} (or in the Appendix A of \cite{KKWZW}).

In this subsection, we study a well-known class of rational RG flows \cite{H82,H83,L15,KKWZW}, $\wf g_k+\phi_\text{adj}\to\wf g_{k'}$.

\subsubsection{$\wf{su}(3)_2+\phi_\text{adj}\to\wf{su}(3)_1$.}\label{su32tosu31}
The UV theory has the MFC
\[ \mc B_\uv\simeq\fib\boxtimes\vecG_{\mbb Z/3\mbb Z}^1=\{1,z\}\boxtimes\{1,x,y\}. \]
The relevant operator $\phi_\text{adj}=\phi_{\widehat{[0;1,1]}}$ preserves the modular fusion subcategory
\[ \mc S_\uv\simeq\vecG_{\mbb Z/3\mbb Z}^1. \]
It flows to the IR subsymmetry
\[ \mc S_\ir\simeq\vecG_{\mbb Z/3\mbb Z}^1. \]
The IR MFC is consistent \cite{KKWZW}, and no emergent symmetry objects are needed:
\[ \mc B_\ir\simeq\vecG_{\mbb Z/3\mbb Z}^1=\{1,X,Y\}. \]
By our notation, the symmetry objects are identified as\footnote{The WZW models are known to have various anomalies \cite{H82,H83,GW86,FO15,NY17,TS18,YHO18,KY19,LHYO22}. Matching the anomalies, we revealed explicit identifications in \cite{KKWZW}. In this example, one identification is
\[ x\leftrightarrow\widehat{[0;2,0]},\quad y\leftrightarrow\widehat{[0;0,2]},\quad X\leftrightarrow\widehat{[0;0,1]},\quad Y\leftrightarrow\widehat{[0;1,0]}. \]
(The other identification is given by swapping $x\leftrightarrow y$ and $X\leftrightarrow Y$, but these are just a change of names.)}
\begin{equation}
\begin{array}{cccc}
\wf{su}(3)_2:&1_0&x_{\frac23}&y_{\frac23}\\
&\downarrow&\downarrow&\downarrow\\
\wf{su}(3)_1:&1_0&X_{\frac13}&Y_{\frac13}
\end{array}.\label{su32tosu31sym}
\end{equation}
The half-integer condition is obeyed:
\[ (h_1,h_{xX},h_{yY})=(0,1,1). \]

Let us identify the category $\mc S_A^0$. While we do not know its categorical dimension, connected étale algebras in $\mc S:=\mc S_\uv\boxtimes\mc S_\ir\simeq\vecG_{\mbb Z/3\mbb Z}^1\boxtimes\vecG_{\mbb Z/3\mbb Z}^1$ have been classified in \cite{KK24rank9}. There are two nontrivial connected étale algebras
\[ A\cong1\oplus xX\oplus yY,1\oplus xY\oplus yX \]
giving the extended vertex operator (super)algebra of an RG theory. Since they are actually Lagrangian algebras, they both give
\[ \mc S_A^0\simeq\vect. \]
Thus, $\mc S_A^0$ is unique up to braided equivalence. This is completely anisotropic. The example gives further support of our conjecture beyond minimal models.

\subsubsection{$(\wf{e_7})_3+\phi_\text{adj}\to(\wf{e_7})_1$.}\label{e73toe71}
The relevant deformation preserves the modular fusion subcategory
\[ \mc S_\uv\simeq\vecG_{\mbb Z/2\mbb Z}^{-1}=\{1,x\}. \]
It flows to the IR subsymmetry
\[ \mc S_\ir\simeq\vecG_{\mbb Z/2\mbb Z}^{-1}=\{1,X\}. \]
The IR MFC is consistent \cite{KKWZW}, and no emergent symmetry objects are necessary. The symmetry objects are identified as
\begin{equation}
\begin{array}{ccc}
(\wf{e_7})_3:&1_0&x_{\frac94}\\
&\downarrow&\downarrow\\
(\wf{e_7})_1:&1_0&X_{\frac34}
\end{array}.\label{e73toe71sym}
\end{equation}
The half-integer condition is obeyed:
\[ (h_1,h_{xX})=(0,3). \]
Let us try to identify $\mc S_A^0$. We do not know its categorical dimension, but $A\in\mc S\simeq\vecG_{\mbb Z/2\mbb Z}^{-1}\boxtimes\vecG_{\mbb Z/2\mbb Z}^{-1}$ have been classified in \cite{KK23rank5}. The only nontrivial one is
\[ A\cong1\oplus xX. \]
It is actually a Lagrangian algebra, and gives
\[ \mc S_A^0\simeq\vect. \]
The category is completely anisotropic. This example further supports our conjecture.

\subsection{$E$-type minimal model $(A_{10},E_6)$}\label{A10E6toM43}
Finally, encouraged by previous tests and armed with our new constraints on rational RG flows, let us try to solve an actual problem. The IR RCFT from the $E$-type minimal model $(A_{10},E_6)$ has remained controversial.

It turns out that the relevant operator with conformal dimension $\frac1{11}$ triggers rational RG flow to $M(4,3)$, the critical Ising model. A reader may be convinced by the scenario if one realizes the UV and IR theories have coset descriptions \cite{GKO86,KW88,K03}
\[ (A_{10},E_6)=\frac{(\wf{e_8})_2\times(\wf{e_8})_1}{(\wf{e_8})_3},\quad M(4,3)=\frac{(\wf{e_8})_1\times(\wf{e_8})_1}{(\wf{e_8})_2}. \]
With the description, the relevant deformation of the UV theory with the adjoint primary falls into the class \cite{R92}. Since the coset descriptions seem not well-known, let us first review the descriptions.

A coset model $\wf g_k/\wf h_l$ has central charge
\[ c(\wf g_k/\wf h_l)=c(\wf g_k)-c(\wf h_l). \]
Thus, the two cosets have central charges
\[ c(\frac{(\wf{e_8})_2\times(\wf{e_8})_1}{(\wf{e_8})_3})=\frac{31}2+8-\frac{248}{11}=\frac{21}{22},\quad c(\frac{(\wf{e_8})_1\times(\wf{e_8})_1}{(\wf{e_8})_2})=2\times8-\frac{31}2=\frac12, \]
matching with $(A_{10},E_6)$ and $M(4,3)$, respectively. For the latter, this uniquely identifies the coset as the critical Ising model. One can compute conformal dimensions of primary operators to convince oneself the former is the $E$-type minimal model and not $M(12,11)$. Here, instead, we give a simpler check which will also be useful later.

The MFCs of the two models $(A_{10},E_6)$ and $M(12,11)$ have different categorical dimensions. They are
\[ D_{(A_{10},E_6)}=\frac{\sqrt{11}}{\sin\frac\pi{11}},\quad D_{M(12,11)}=\frac{\sqrt{33}(1+\sqrt3)}{\sin\frac\pi{11}}. \]
These are given by the inverse of the modular $S$-matrix element $S_{11}$. Since MFCs of $E_8$ WZW models are identified as
\[ \mc B_{(\wf{e_8})_1}\simeq\vect,\quad\mc B_{(\wf{e_8})_2}\simeq\ising,\quad\mc B_{(\wf{e_8})_3}\simeq psu(2)_9, \]
the first coset has
\[ S_{11}=1\times\frac12\times\frac1{\sqrt{\frac{11}{4\sin^2\frac\pi{11}}}}=\frac1{\sqrt{\frac{11}{\sin^2\frac\pi{11}}}}, \]
or
\[ D=\frac{\sqrt{11}}{\sin\frac\pi{11}}. \]
Note that since $(\wf{e_8})_1$ is trivial, there is no identification of coset fields, and hence there is no factor of 2 unlike $M(m+1,m)$. The result uniquely identifies the coset as the $E$-type minimal model $(A_{10},E_6)$.

According to \cite{R92}, the relevant deformation of the diagonal coset $\frac{(\wf{e_8})_2\times(\wf{e_8})_1}{(\wf{e_8})_3}$ with the adjoint primary triggers a rational RG flow to another diagonal coset $\frac{(\wf{e_8})_1\times(\wf{e_8})_1}{(\wf{e_8})_2}$. The adjoint primary corresponds to the affine weight $\widehat{[1;0,0,0,0,0,0,1,0]}$ in $(\wf{e_8})_3$. It has conformal dimension
\[ h_{[1,1;\text{adj}]}=h^2_{\widehat{[2;0,0,0,0,0,0,0,0]}}+h^1_{\widehat{[1;0,0,0,0,0,0,0,0]}}-h^3_{\widehat{[1;0,0,0,0,0,0,1,0]}}=-\frac{30}{3+30}=\frac1{11}\quad(\mods1). \]
The relevant operator preserves modular fusion subcategory
\[ \mc S_\uv\simeq\ising \]
of UV MFC
\[ \mc B_\uv\simeq\ising\boxtimes psu(2)_9=\{1,x,y\}\boxtimes\{1,s,t,v,w\}. \]
The surviving symmetries flow to the IR subsymmetry
\[ \mc S_\ir\simeq\ising=\{1,X,Y\}. \]
Note that the IR symmetry category is consistent, and no emergent symmetry objects are needed. Indeed, the scenario
\begin{equation}
\begin{array}{cccc}
(A_{10},E_6):&1_0&x_{\frac72}&y_{\frac{31}{16}}\\
&\downarrow&\downarrow&\downarrow\\
M(4,3):&1_0&X_{\frac12}&Y_{\frac1{16}}
\end{array}\label{A10E6toM43sym}
\end{equation}
obeys the half-integer condition:
\[ (h_1,h_{xX},h_{yY})=(0,4,2). \]

Since this is where disagreement could occur, let us make a few comments. We first note that a possible IR theory $M(5,4)$ does \textit{not} obey the half-integer condition for the non-invertible simple object $y$:
\[ \frac{31}{16}+\frac7{16}\notin\frac12\mbb Z. \]
We cannot rule out this scenario because we cannot rule out an existence of rational RG flows without RG domain wall/interface or boundary condition. What we do know is that if the flow to $M(5,4)$ existed, it could not be described by RG domain wall/interface or boundary condition. We also note that the scenario is unnatural because it requires mysterious lift of free energy; since the MFC $\ising$ is already consistent, the emergent symmetry objects demands explanation. At the moment, we do not know any.

The observation that $(A_{10},E_6)\to M(5,4)$ does not satisfy the half-integer condition has an interesting implication. In \cite{NK22}, with some assumptions, Nakayama and the author narrowed down possible UV theories flowing to fermionic $M(5,4)$ minimal model \cite{P88,HNT20} to two, either fermionic $M(11,4)$ minimal model or fermionic $(A_{10},E_6)$ minimal model \cite{K20}. If we assume the rational RG flow admit RG domain wall/interface or boundary condition, we can rule out the latter. Namely, the UV theory leading to supersymmetric Gross-Neveu-Yukawa fixed point is the fermionic $M(11,4)$ minimal model.

After these comments, let us go back to our scenario. In order to test our conjecture, let us identify the category $\mc S_A^0$. In this case, an RG theory is known:
\[ \frac{(\wf{e_8})_1\times(\wf{e_8})_1\times(\wf{e_8})_1}{(\wf{e_8})_3}. \]
As the numerator is trivial, its MFC is simply given by the denominator. Thus, the MFC of an RG theory has categorical dimension
\[ D(\mc B_\rg)=\frac1{S_{11}}=\sqrt{\frac{11}{4\sin^2\frac\pi{11}}}. \]
By removing contributions from the broken $psu(2)_9$, we find
\[ D(\mc S_A^0)=1, \]
or
\[ \fp(\mc S_A^0)=1. \]
Our conjecture claims this is the category of dyslectic (right) $A$-modules for a maximal connected étale algebra $A\in\mc S:=\mc S_\uv\boxtimes\mc S_\ir\simeq\ising\boxtimes\ising$. Connected étale algebras in the rank nine MFC have been classified in \cite{KK24rank9}. For our MFC, the maximal one is
\[ A\cong1\oplus xX\oplus yY. \]
It is a Lagrangian algebra giving
\[ \mc S_A^0\simeq\vect. \]
The identification matches the Frobenius-Perron dimension. The category is completely anisotropic, and gives further support of our conjecture.

In order to see the power of the half-integer condition and to give another support of the flow, let us pretend we do not know the conjecture \cite{R92} but assume the flow admits RG domain wall/interface. The surviving modular fusion category $\mc S_\uv\simeq\ising$ flows to $\mc S_\ir\simeq\ising$.

The $\mbb Z/2\mbb Z$ simple object in $\ising$ always has conformal dimension $\frac12$ mod 1. Thus, it trivially satisfies the half-integer condition. Let us look at the non-invertible simple object. It can have conformal dimensions $\frac{2n+1}{16}$ with $n=0,1,\dots,7$. In UV, it has $h_y=\frac{31}{16}$. For its image $Y=F(y)$ to satisfy the half-integer condition, the IR conformal dimension should be either $\frac1{16}$ or $\frac9{16}$. They give MFCs with additive central charges $\frac12,-\frac72$ mod 8, respectively. For the latter to be unitary, the smallest additive central charge is $\frac92$, which is larger than UV. This rules out the latter possibility as follows. All CFTs with central charge $0<c<1$ have been classified \cite{KL02}, and they are minimal models. None of minimal models has $\ising$ with $h=\frac9{16}$ for the non-invertible simple object, and central charge smaller than $c_\uv=\frac{21}{22}$. Therefore, the only possibility is $\ising$ with $h=\frac1{16}$ for the non-invertible simple object. The only possibility is the critical Ising model $M(4,3)$.

In fact, we can do a little more; give conjectures on signs of the relevant coupling with the Cardy's method \cite{Cardy17}. With the Lagrangian coupling $\delta S\ni\lambda_{1/11}\int O_{1/11}$, we try to minimize the energy
\[ E_a=\frac{\pi\cdot21/22}{24(2\tau_1)^2}-\frac{S_{ja}}{S_{1a}}\frac{\lambda_{1/11}}{(2\tau_a)^{1/11}} \]
for a Cardy state $|a\ra$ with variational parameter $\tau_a$. Here, $j$ is the index corresponding to the relevant operator $O_{1/11}$. The ratio of the modular $S$-matrices take the largest value ($\approx3.228$) for Cardy states with conformal dimensions $0,\frac{31}{16},\frac72$, and the smallest value ($\approx-1.398$) for those with $\frac1{11},\frac{133}{176},\frac7{22}$. Therefore, for positive Lagrangian coupling, the first three Cardy states give (approximate) vacuum, and for negative Lagrangian coupling, the second three Cardy states give threefold degenerate vacua. Since the first case contains the identity, this signals the massless flow. For the second case, the ground state degeneracy (GSD) three is consistent with mathematical results \cite{KK23GSD,KK23preMFC,KK24rank9}. To summarize, we arrive the conjecture
\begin{equation}
    (A_{10},E_6)+\lambda_{1/11}O_{1/11}=\begin{cases}M(4,3)&(\lambda_{1/11}>0),\\\text{TQFT w/ GSD}=3&(\lambda_{1/11}<0).\end{cases}\label{signs}
\end{equation}

\section*{Acknowledgment}
We thank Yuto Moriwaki for discussions and teaching us many important references. We also thank Yuto Moriwaki and Yu Nakayama for comments on the draft.

\appendix
\setcounter{section}{0}
\renewcommand{\thesection}{\Alph{section}}
\setcounter{equation}{0}
\renewcommand{\theequation}{\Alph{section}.\arabic{equation}}

\section{Preliminary}\label{preliminary}
In this appendix, we collect relevant definitions and facts. It is intended for physics-oriented readers; we refer more details to standard textbooks, say, \cite{EGNO15}.

\subsection{Category}
Let $\mc C$ be a fusion category over the complex field $\mbb C$. One of the simplest ways to understand a fusion category is its analogy with representations. Physicists know the two-dimensional irreducible representation (irrep) 2 of $SU(2)$ obeys
\[ 2\otimes2=4=1\oplus3. \]
We see the representations are equipped with two binary operations, tensor product and direct sum. A fusion category $\mc C$ is an analogue of this; it is a collection of simple object (corresponding to irreps) equipped with fusion (or monoidal) product (corresponding to tensor product) and direct sum. Since the fusion product forms a monoid, a fusion category contains the identity object $1\in\mc C$ which is simple. The identity object obeys
\begin{equation}
    \forall c\in\mc C,\quad 1\otimes c\cong c\cong c\otimes1.\label{identityobject}
\end{equation}
We also see generic object corresponding to reducible representation (such as 4) can be written as a direct sum of finitely many simple objects (semisimplicity). However, unlike usual representations, a fusion category is finite by definition, and has only finitely many simple objects. The number of simple objects in $\mc C$ is called the rank of the fusion category and denoted $\rank(\mc C)$. The simple objects are denoted $c_i$ with $i=1,2,\dots,\rank(\mc C)$. The fusion product is characterized by fusion coefficients ${N_{ij}}^k\in\mbb N$:
\[ c_i\otimes c_j\cong\bigoplus_{k=1}^{\rank(\mc C)}{N_{ij}}^kc_k. \]
The fusion coefficients define $\mbb N$-matrices
\[ (N_i)_{jk}:={N_{ij}}^k. \]
Since the matrix elements are non-negative, the Perron-Frobenius theorem guarantees an existence of positive largest eigenvalue $\fp_\mc C(c_i)$ of $N_i$ called the Frobenius-Perron dimension of $c_i\in\mc C$. The Frobenius-Perron dimension of the category $\mc C$ is defined by their squared sum:
\[ \fp(\mc C):=\sum_{j=1}^{\rank(\mc C)}(\fp_\mc C(c_j))^2. \]
In a spherical fusion category, we can define quantum dimension $d_i$ of $c_i$. They obey the same multiplication rule
\[ d_id_j=\sum_{k=1}^{\rank(\mc C)}{N_{ij}}^kd_k. \]
The categorical (or global) dimension of $\mc C$ is defined by their squared sum
\[ D(\mc C)^2:=\sum_{j=1}^{\rank(\mc C)}d_j^2. \]
A spherical fusion category is called pseudo-unitary if $D(\mc C)^2=\fp(\mc C)$. A spherical fusion category is called unitary if $\forall c_i\in\mc C$, $d_i=\fp_\mc C(c_i)$.

A braided fusion category (BFC) is a pair $(\mc B,c)$ of a fusion category and a braiding $c$. (In order to avoid confusion, we denote BFCs by $\mc B$, and its simple objects by $b_i$'s. We often abbreviate a pair by $\mc B$.) A braiding is a natural isomorphism
\[
\begin{split}
    c:-\otimes-&\stackrel\sim\Rightarrow-\otimes-\\
    c_{b_i,b_j}:b_i\otimes b_j&\cong b_j\otimes b_i
\end{split},
\]
obeying consistency conditions (called hexagon axiom). A reverse braiding $c^\text{rev}$ of $c$ is defined by
\[ c^\text{rev}_{b_i,b_j}:=c_{b_j,b_i}^{-1}. \]
A BFC $\mc B^\text{rev}=(\mc B,c^\text{rev})$ is called a reverse BFC. A braiding is characterized by conformal dimensions $h_i$ of $c_i$; the topological twist of $b_i$ is defined by $\theta_i:=e^{2\pi ih_i}$. Objects with trivial topological twists $\theta=1$ are called bosons. One notices that if one performs braiding twice, objects go back to the original order. It is characterized by double braiding
\begin{equation}
    c_{b_j,b_i}\circ c_{b_i,b_j}\cong\bigoplus_{k=1}^{\rank(\mc B)}{N_{ij}}^k\frac{e^{2\pi ih_k}}{e^{2\pi i(h_i+h_j)}}id_k,\label{doublebraiding}
\end{equation}
where $id_k$ is the identity morphism at $b_k$. A simple object $b_i$ is called transparent if
\begin{equation}
    \forall b_j\in\mc B,\quad c_{b_j,b_i}\circ c_{b_i,b_j}\cong id_{b_i\otimes b_j}.\label{transparent}
\end{equation}
By (\ref{identityobject}), the identity object is transparent. The collection $\mc B'$ of transparent simple objects is called the symmetric (or M\"uger) center:
\begin{equation}
    \mc B':=\{b_i\in\mc B|\forall b_j\in\mc B,\ c_{b_j,b_i}\circ c_{b_i,b_j}\cong id_{b_i\otimes b_j}\}.\label{symmetriccenter}
\end{equation}
A BFC is called symmetric if all simple objects are transparent. On the other extreme, a BFC is called non-degenerate if the identity object is the only transparent object, $\mc B'=\{1\}$. A non-degenerate BFC $\mc B\neq\vect$ is called prime if there is no proper non-degenerate braided fusion subcategory other than $\vect$. A generic non-degenerate BFC $\mc B\neq\vect$ is a Deligne tensor product of prime non-degenerate braided fusion subcategories \cite{DMNO10}
\[ \mathcal B\simeq\mc B_1\boxtimes\mc B_2\boxtimes\cdots\boxtimes\mc B_n. \]

A spherical BFC is called pre-modular fusion category (pre-MFC). The Gauss sum of a pre-MFC $\mc B$ is defined by
\[ \tau^\pm(\mc B):=\sum_{j=1}^{\rank(\mc B)}\theta_j^{\pm1}d_j^2. \]
It defines multiplicative central charge
\[ \xi(\mc B):=\frac{\tau^+(\mc B)}{D(\mc B)}. \]
The familiar additive central charge $c(\mc B)$ (mod 8) is defined by
\[ \xi(\mc B)=e^{2\pi ic(\mc B)/8}. \]
The quantum trace of the double braiding (\ref{doublebraiding}) defines (unnormalized) modular $S$-matrix
\begin{equation}
    \widetilde S_{ij}:=\tr(c_{b_j,b_i}\circ c_{b_i,b_j})=\sum_{k=1}^{\rank(\mc B)}{N_{ij}}^k\frac{e^{2\pi ih_k}}{e^{2\pi i(h_i+h_j)}}d_k.\label{unnormalizedS}
\end{equation}
The non-degeneracy of a pre-MFC is equivalent to the pre-MFC with non-singular modular $S$-matrix. A non-degenerate pre-MFC is called modular or modular fusion category (MFC). The normalized modular $S$-matrix is defined by
\begin{equation}
    S_{ij}:=\frac{\widetilde S_{ij}}{D(\mc B)}.\label{normalizedS}
\end{equation}

The Drinfeld center (or quantum double) $\mc Z(\mc C)$ of a fusion category $\mc C$ consists of pairs $(c,\gamma_c)$ of an object $c\in\mc C$ and a natural isomorphism $\gamma_c:c'\otimes c\cong c\otimes c'$ for $c'\in\mc C$ \cite{M01,DGNO09}. They are subject to consistency conditions. It has a forgetful functor $U:\mc Z(\mc C)\to\mc C$ sending $(c,\gamma_c)\mapsto c$. The Drinfeld center is a non-degenerate BFC \cite{M01,DGNO09,T16}.

\subsection{Functor}
Let $\mc C_{1,2}$ be fusion categories. A monoidal functor is a pair $(F,J)$ of a functor $F:\mc C_1\to\mc C_2$ and a natural isomorphism $J_{c_i,c_j}:Fc_i\otimes_2Fc_j\cong F(c_i\otimes_1c_j)$ subject to consistency conditions. We often denote a monoidal functor by $F$.

Let $(\mc B_1,c^1),(\mc B_2,c^2)$ be BFCs. A braided (monoidal) functor $F:\mc B_1\to\mc B_2$ is a monoidal functor $(F,J)$ obeying
\[ \forall b_i,b_j\in\mc B_1,\quad F(c^1_{b_i,b_j})\cdot J_{b_i,b_j}=J_{b_j,b_i}\cdot c^2_{Fb_i,Fb_j}. \]
Let $\mc B$ be a BFC and $\mc C$ a fusion category. A central functor $F:\mc B\to\mc C$ is a monoidal functor $(F,J):\mc B\to\mc C$ admitting a braided monoidal functor $F':\mc B\to\mc Z(\mc C)$ such that $U\cdot F'=F$.

Let $\mc E$ be a symmetric fusion category. A fusion category over $\mc E$ is a fusion category $\mc C$ with a central functor $F:\mc E\to\mc C$. A BFC over $\mc E$ (or braided fusion $\mc E$-category) is a BFC $\mc B$ with a braided monoidal functor $F:\mc E\to\mc B'$ \cite{DGNO09}. Let $\mc B$ be a BFC over $\mc E$. $\mc B$ is called non-degenerate over $\mc E$ if $\mc B'=\mc E$ \cite{DNO11}. (Thus, by definition, any BFC $\mc B$ is non-degenerate over its center $\mc B'$.) Let $\mc C$ be a fusion category over $\mc E$ with a fully faithful central functor. The centralizer of $\mc E$ in $\mc Z(\mc C)$ is defined by
\[ \mc Z(\mc C,\mc E):=\{(c,\gamma_c)\in\mc Z(\mc C)|\forall e\in\mc E,\ c_{e,(c,\gamma_c)}\circ c_{(c,\gamma_c),e}=id_{(c,\gamma_c)\otimes e}\}. \]

\subsection{Algebra}
An algebra in a fusion category $\mc C$ is a monoid. Namely, it is a triple $(A,\mu,\eta)$ of object $A\in\mc C$, and two morphisms $\mu:A\otimes A\to A$ and $\eta:1\to A$ obeying consistency conditions. (We often denote an algebra by $A$.) An algebra is called connected if $\dim_\mbb C\mc C(1,A)=1$. Let $\mc C$ be a fusion category and $(A,\mu,\eta)$ an algebra. A right $A$-module is a pair $(m,p)$ of an object $m\in C$ and a morphism $p:m\otimes A\to m$ subject to consistency conditions. They form the category $\mc C_A$ of right $A$-modules. An algebra $A$ is called separable if $\mc C_A$ is separable.

Let $\mc B$ be a BFC. An algebra $(A,\mu,\eta)$ in a BFC $(\mc B,c)$ is called commutative if $\mu=\mu\cdot c_{A,A}$. An étale algebra is a commutative separable algebra. A connected étale algebra $A$ in a non-degenerate BFC $\mc B$ is called a Lagrangian algebra if $(\fp_\mc B(A))^2=\fp(\mc B)$. Let $A\in\mc B$ be an algebra. The category $\mc B_A$ of right $A$-modules has an important subcategory $\mc B_A^0\subset\mc B_A$. It consists of dyslectic (or local) right $A$-modules \cite{P95} $(m,p)\in\mc B_A$ obeying
\[ p\cdot c_{A,m}\cdot c_{m,A}=p. \]
For a conencted étale algebra $A$ in a non-degenerate BFC $\mc B$, the categories obey \cite{BE99,BEK99,BEK00,ENO02,KO01,DMNO10}
\begin{equation}
\begin{split}
    \fp(\mc B_A)&=\frac{\fp(\mc B)}{\fp_\mc B(A)},\\
    \fp(\mc B_A^0)&=\frac{\fp(\mc B)}{(\fp_\mc B(A))^2},\\
    c(\mc B)&=c(\mc B_A^0)\quad\mods8.
\end{split}\label{FPdimformula}
\end{equation}
The identity object $1$ always gives the trivial connected étale algebra. Non-degenerate BFCs without nontrivial connected étale algebras are called completely anisotropic. A BFC $\mc B$ over a symmetric fusion category $\mc E$ is called completely $\mc E$-anisotropic if all connected étale algebras in $\mc B$ belong to $\mc E$.

Let $\mc B_{1,2}$ be non-degenerate BFCs. They are called Witt equivalent if there exist fusion categories $\mc C_{1,2}$ such that $\mc B_1\boxtimes\mc Z(\mc C_1)\simeq\mc B_2\boxtimes\mc Z(\mc C_2)$ is a braided equivalence \cite{DMNO10}. This introduces equivalence classes to non-degenerate BFCs. The Witt equivalence class of $\mc B$ is denoted by $[\mc B]$. For a non-degenerate BFC $\mc B$ and a connected étale algebra $A\in\mc B$, we have \cite{DMNO10}
\begin{equation}
    [\mc B]=[\mc B_A^0].\label{BBA0Wittequivalent}
\end{equation}

Let $\mc B_{1,2}$ be non-degenerate BFCs over a symmetric fusion category $\mc E$ such that the corresponding braided monoidal functors $\mc E\to\mc B_i\ (i=1,2)$ are fully faithful. The $\mc B_{1,2}$ are called Witt equivalent if there exist fusion categories $\mc C_{1,2}$ over $\mc E$ such that $\mc B_1\boxtimes_\mc E\mc Z(\mc C_1,\mc E)\simeq\mc B_2\boxtimes_\mc E\mc Z(\mc C_2,\mc E)$ is a braided equivalence. It introduces an equivalence relation to non-degenerate BFCs over $\mc E$. The Witt equivalence class of $\mc B$ is denoted $[\mc B]$. The set of Witt equivalence classes of non-degenerate BFCs over $\mc E$ is denoted by $\mc W(\mc E)$. For a non-degenerate BFC $\mc B$ over $\mc E$ and a connected étale algebra such that $A\cap\mc E=1$, i.e., $A$ does not contain simple objects in $\mc E$ except the identity object, we have \cite{DNO11}
\begin{equation}
    [\mc B]=[\mc B_A^0].\label{BBA0overEWittequivalent}
\end{equation}
By Theorem 5.5 of \cite{DNO11}, for all Witt equivalence classes, there exists a unique (up to braided equivalence) completely $\mc E$-anisotropic representative.


\begin{thebibliography}{30}
\bibitem{tH79}
  G.~'t Hooft,
  ``Naturalness, chiral symmetry, and spontaneous chiral symmetry breaking,''
  NATO Sci. Ser. B \textbf{59}, 135-157 (1980)
  doi:10.1007/978-1-4684-7571-5\_9
\bibitem{GKSW14}
  D.~Gaiotto, A.~Kapustin, N.~Seiberg and B.~Willett,
  ``Generalized Global Symmetries,''
  JHEP \textbf{02}, 172 (2015)
  doi:10.1007/JHEP02(2015)172
  [arXiv:1412.5148 [hep-th]].
\bibitem{V88}
  E.~P.~Verlinde,
  ``Fusion Rules and Modular Transformations in 2D Conformal Field Theory,''
  Nucl. Phys. B \textbf{300}, 360-376 (1988) doi:10.1016/0550-3213(88)90603-7
\bibitem{MS88}
  G.~W.~Moore and N.~Seiberg,
  ``Classical and Quantum Conformal Field Theory,'' Commun. Math. Phys. \textbf{123}, 177 (1989) doi:10.1007/BF01238857
\bibitem{MS89}
  G.~W.~Moore and N.~Seiberg,
  ``LECTURES ON RCFT,'' RU-89-32.
\bibitem{PZ00}
  V.~B.~Petkova and J.~B.~Zuber,
  ``Generalized twisted partition functions,''
  Phys. Lett. B \textbf{504}, 157-164 (2001)
  doi:10.1016/S0370-2693(01)00276-3 [arXiv:hep-th/0011021 [hep-th]].
\bibitem{FRS02}
  J.~Fuchs, I.~Runkel and C.~Schweigert,
  ``TFT construction of RCFT correlators 1. Partition functions,''
  Nucl. Phys. B \textbf{646}, 353-497 (2002)
  doi:10.1016/S0550-3213(02)00744-7
  [arXiv:hep-th/0204148 [hep-th]].
\bibitem{FRS03}
  J.~Fuchs, I.~Runkel and C.~Schweigert,
  ``TFT construction of RCFT correlators. 2. Unoriented world sheets,''
  Nucl. Phys. B \textbf{678}, 511-637 (2004)
  doi:10.1016/j.nuclphysb.2003.11.026
  [arXiv:hep-th/0306164 [hep-th]].
\bibitem{FRS04a}
  J.~Fuchs, I.~Runkel and C.~Schweigert,
  ``TFT construction of RCFT correlators. 3. Simple currents,''
  Nucl. Phys. B \textbf{694}, 277-353 (2004)
  doi:10.1016/j.nuclphysb.2004.05.014
  [arXiv:hep-th/0403157 [hep-th]].
\bibitem{FRS04b}
  J.~Fuchs, I.~Runkel and C.~Schweigert,
  ``TFT construction of RCFT correlators IV: Structure constants and correlation functions,''
  Nucl. Phys. B \textbf{715}, 539-638 (2005)
  doi:10.1016/j.nuclphysb.2005.03.018
  [arXiv:hep-th/0412290 [hep-th]].
\bibitem{FGS09}
  S.~Fredenhagen, M.~R.~Gaberdiel and C.~Schmidt-Colinet,
  ``Bulk flows in Virasoro minimal models with boundaries,''
  J. Phys. A \textbf{42}, no.49, 495403 (2009)
  doi:10.1088/1751-8113/42/49/495403
  [arXiv:0907.2560 [hep-th]].
\bibitem{G12}
  D.~Gaiotto,
  ``Domain Walls for Two-Dimensional Renormalization Group Flows,''
  JHEP \textbf{12}, 103 (2012)
  doi:10.1007/JHEP12(2012)103
  [arXiv:1201.0767 [hep-th]].
\bibitem{BT17}
  L.~Bhardwaj and Y.~Tachikawa,
  ``On finite symmetries and their gauging in two dimensions,''
  JHEP \textbf{03}, 189 (2018)
  doi:10.1007/JHEP03(2018)189
  [arXiv:1704.02330 [hep-th]].
\bibitem{KKARG}
  K.~Kikuchi,
  ``Axiomatic rational RG flow,''
  [arXiv:2209.00016 [hep-th]].
\bibitem{KK21}
  K.~Kikuchi,
  ``Symmetry enhancement in RCFT,''
  [arXiv:2109.02672 [hep-th]].
\bibitem{KKSUSY}
  K.~Kikuchi,
  ``Emergent SUSY in two dimensions,''
  [arXiv:2204.03247 [hep-th]].
\bibitem{KK22II}
  K.~Kikuchi,
  ``Symmetry enhancement in RCFT II,''
  [arXiv:2207.06433 [hep-th]].
\bibitem{KK22free}
  K.~Kikuchi,
  ``Emergent symmetry and free energy,''
  [arXiv:2207.10095 [hep-th]].
\bibitem{NK22}
  Y.~Nakayama and K.~Kikuchi,
  ``The fate of non-supersymmetric Gross-Neveu-Yukawa fixed point in two dimensions,''
  [arXiv:2212.06342 [hep-th]].
\bibitem{KKWZW}
  K.~Kikuchi,
  ``RG flows from WZW models,''
  [arXiv:2212.13851 [hep-th]].
\bibitem{TN24}
  T.~Tanaka and Y.~Nakayama,
  ``Infinitely many new renormalization group flows between Virasoro minimal models from non-invertible symmetries,''
  [arXiv:2407.21353 [hep-th]].
\bibitem{CKM17}
  T.~Creutzig, S.~Kanade and R.~McRae, ``Tensor categories for vertex operator superalgebra extensions,'' Mem. Amer. Math. Soc. 295 (2024), no. 1472
  doi:10.1090/memo/1472
  [arXiv:1705.05017 [math.QA]].
\bibitem{KO01}
  A.~Kirillov Jr. and V.~Ostrik, ``ON A q-ANALOG OF THE MCKAY CORRESPONDENCE AND THE ADE CLASSIFICATION OF slb2 CONFORMAL FIELD,'' Advances in Mathematics 171(2002), 183-227. https://doi.org/10.1006/aima.2002.2072 [arXiv:math/0101219 [math.QA]].
\bibitem{KL02}
  Y.~Kawahigashi and R.~Longo,
  ``Classification of local conformal nets: Case c \ensuremath{<} 1,''
  Annals Math. \textbf{160}, 493-522 (2004)
  [arXiv:math-ph/0201015 [math-ph]].
\bibitem{EP09}
   D.~E.~Evans and M.~Pugh, ``$SU(3)$-Goodman-de la Harpe-Jones subfactors and the realization of $SU(3)$ modular invariants,'' Rev. Math. Phys. 21 (2009), 877–928. https://doi.org/10.1142/S0129055X09003761
   [arXiv:0906.4252 [math.OA]].
\bibitem{DMNO10}
  A.~Davydov, M.~Müger, D.~Nikshych and V.~Ostrik, ``The Witt group of non-degenerate braided fusion categories,'' Journal für die reine und angewandte Mathematik (Crelles Journal), 2013(677), 135-177. https://doi.org/10.1515/crelle.2012.014
  [arXiv:1009.2117 [math.QA]].
\bibitem{DNO11}
  A.~Davydov, D.~Nikshych and V.~Ostrik, ``On the structure of the Witt group of braided fusion categories,'' Sel. Math. New Ser. 19, 237–269 (2013). https://doi.org/10.1007/s00029-012-0093-3
  [arXiv:1109.5558 [math.QA]].
\bibitem{EM21}
  C.~Edie-Michell, ``TYPE II QUANTUM SUBGROUPS OF slN . I: SYMMETRIES OF LOCAL MODULES,'' Communications of the American Mathematical Society 3(2023), 112-165. https://doi.org/10.1090/cams/19
  [arXiv:2102.09065 [math.QA]].
\bibitem{CEM23}
  D.~Copeland and C.~Edie-Michell, ``CELL SYSTEMS FOR Rep(Uq(slN )) MODULE CATEGORIES,''
  [arXiv:2301.13172 [math.QA]].
\bibitem{G23}
  T.~Gannon,
  ``Exotic quantum subgroups and extensions of affine Lie algebra VOAs -- part I,''
  [arXiv:2301.07287 [math.QA]].
\bibitem{KK23GSD}
  K.~Kikuchi,
  ``Ground state degeneracy and module category,''
  [arXiv:2311.00746 [hep-th]].
\bibitem{KK23preMFC}
  K.~Kikuchi,
  ``Classification of connected \'etale algebras in pre-modular fusion categories up to rank three,''
  [arXiv:2311.15631 [math.QA]].
\bibitem{KK23rank5}
  K.~Kikuchi,
  ``Classification of connected \'etale algebras in modular fusion categories up to rank five,''
  [arXiv:2312.13353 [math.QA]].
\bibitem{KKH24}
  K.~Kikuchi, K.~S.~Kam and F.~H.~Huang,
  ``Classification of connected \'etale algebras in multiplicity-free modular fusion categories at rank six,''
  [arXiv:2402.00403 [math.QA]].
\bibitem{KK24rank9}
  K.~Kikuchi,
  ``Classification of connected \'etale algebras in multiplicity-free modular fusion categories up to rank nine,''
  [arXiv:2404.16125 [math.QA]].
\bibitem{HKL14}
  Y.~Z.~Huang, A.~Kirillov and J.~Lepowsky,
  ``Braided tensor categories and extensions of vertex operator algebras,''
  Commun. Math. Phys. \textbf{337}, no.3, 1143-1159 (2015)
  doi:10.1007/s00220-015-2292-1
  [arXiv:1406.3420 [math.QA]].
\bibitem{M21}
  Y.~Moriwaki,
  ``Quantum coordinate ring in WZW model and affine vertex algebra extensions,''
  Selecta Math. \textbf{28}, no.4, 68 (2022)
  doi:10.1007/s00029-022-00782-2
  [arXiv:2111.11357 [math.QA]].
\bibitem{KP05}
  A.~Kitaev and J.~Preskill,
  ``Topological entanglement entropy,''
  Phys. Rev. Lett. \textbf{96}, 110404 (2006)
  doi:10.1103/PhysRevLett.96.110404
  [arXiv:hep-th/0510092 [hep-th]].
\bibitem{LW05}
  M.~Levin and X.~G.~Wen,
  ``Detecting Topological Order in a Ground State Wave Function,''
  Phys. Rev. Lett. \textbf{96}, 110405 (2006)
  doi:10.1103/PhysRevLett.96.110405
  [arXiv:cond-mat/0510613 [cond-mat.str-el]].
\bibitem{ENO02}
  P.~Etingof, D.~Nikshych and V.~Ostrik,
  ``On fusion categories,''
  Annals of Mathematics 162, no. 2 (2005): 581–642. http://www.jstor.org/stable/20159926
  [arXiv:math/0203060 [math.QA]].
\bibitem{BBR24}
  M.~Balasubramanian, M.~Buican and R.~Radhakrishnan,
  ``On the Classification of Bosonic and Fermionic One-Form Symmetries in $2+1$d and 't Hooft Anomaly Matching,''
  [arXiv:2408.00866 [hep-th]].
\bibitem{cthm}
  A.~B.~Zamolodchikov,
  ``Irreversibility of the Flux of the Renormalization Group in a 2D Field Theory,'' JETP Lett. \textbf{43}, 730-732 (1986)
\bibitem{CDR17}
  O.~A.~Castro-Alvaredo, B.~Doyon and F.~Ravanini,
  ``Irreversibility of the renormalization group flow in non-unitary quantum field theory,''
  J. Phys. A \textbf{50}, no.42, 424002 (2017)
  doi:10.1088/1751-8121/aa8a10
  [arXiv:1706.01871 [hep-th]].
\bibitem{BD11}
  T.~Booker and A.~Davydov, ``Commutative Algebras in Fibonacci Categories,'' Journal of Algebra 355(2012), 176-204. https://doi.org/10.1016/j.jalgebra.2011.12.029
  [arXiv:1103.3537 [math.CT]].
\bibitem{BSS02}
  F.~A.~Bais, B.~J.~Schroers and J.~K.~Slingerland,
  ``Broken quantum symmetry and confinement phases in planar physics,''
  Phys. Rev. Lett. \textbf{89}, 181601 (2002)
  doi:10.1103/PhysRevLett.89.181601
  [arXiv:hep-th/0205117 [hep-th]].
\bibitem{BSS02'}
  F.~A.~Bais, B.~J.~Schroers and J.~K.~Slingerland,
  ``Hopf symmetry breaking and confinement in (2+1)-dimensional gauge theory,''
  JHEP \textbf{05}, 068 (2003)
  doi:10.1088/1126-6708/2003/05/068
  [arXiv:hep-th/0205114 [hep-th]].
\bibitem{BS08}
  F.~A.~Bais and J.~K.~Slingerland,
  ``Condensate induced transitions between topologically ordered phases,''
  Phys. Rev. B \textbf{79}, 045316 (2009)
 doi:10.1103/PhysRevB.79.045316
  [arXiv:0808.0627 [cond-mat.mes-hall]].
\bibitem{K13}
  L.~Kong,
  ``Anyon condensation and tensor categories,''
  Nucl. Phys. B \textbf{886}, 436-482 (2014)
  doi:10.1016/j.nuclphysb.2014.07.003
  [arXiv:1307.8244 [cond-mat.str-el]].
\bibitem{Z90}
  A.~B.~Zamolodchikov,
  ``Thermodynamic Bethe Ansatz in Relativistic Models. Scaling Three State Potts and Lee-yang Models,''
  Nucl. Phys. B \textbf{342}, 695-720 (1990)
  doi:10.1016/0550-3213(90)90333-9
\bibitem{Z91}
  A.~B.~Zamolodchikov,
  ``From tricritical Ising to critical Ising by thermodynamic Bethe ansatz,''
  Nucl. Phys. B \textbf{358}, 524-546 (1991)
  doi:10.1016/0550-3213(91)90423-U
\bibitem{M91}
  M.~J.~Martins,
  ``The Thermodynamic Bethe ansatz for deformed W A(N)-1 conformal field theories,''
  Phys. Lett. B \textbf{277}, 301-305 (1992)
  doi:10.1016/0370-2693(92)90750-X
  [arXiv:hep-th/9201032 [hep-th]].
\bibitem{RST94}
  F.~Ravanini, M.~Stanishkov and R.~Tateo,
  ``Integrable perturbations of CFT with complex parameter: The M(3/5) model and its generalizations,''
  Int. J. Mod. Phys. A \textbf{11}, 677-698 (1996)
  doi:10.1142/S0217751X96000304
  [arXiv:hep-th/9411085 [hep-th]].
\bibitem{DDT00}
  P.~Dorey, C.~Dunning and R.~Tateo,
  ``New families of flows between two-dimensional conformal field theories,''
  Nucl. Phys. B \textbf{578}, 699-727 (2000)
  doi:10.1016/S0550-3213(00)00185-1
  [arXiv:hep-th/0001185 [hep-th]].
\bibitem{FMS}
  P.~Di Francesco, P.~Mathieu and D.~Senechal,
  ``Conformal Field Theory,''
  doi:10.1007/978-1-4612-2256-9
\bibitem{H82}
  F.~D.~M.~Haldane,
  ``Continuum dynamics of the 1-D Heisenberg antiferromagnetic identification with the O(3) nonlinear sigma model,''
  Phys. Lett. A \textbf{93}, 464-468 (1983)
  doi:10.1016/0375-9601(83)90631-X
\bibitem{H83}
  F.~D.~M.~Haldane,
  ``Nonlinear field theory of large spin Heisenberg antiferromagnets. Semiclassically quantized solitons of the one-dimensional easy Axis Neel state,''
  Phys. Rev. Lett. \textbf{50}, 1153-1156 (1983)
  doi:10.1103/PhysRevLett.50.1153
\bibitem{L15}
  P.~Lecheminant,
  ``Massless renormalization group flow in SU(N)$_k$ perturbed conformal field theory,''
  Nucl. Phys. B \textbf{901}, 510-525 (2015)
  doi:10.1016/j.nuclphysb.2015.11.004
  [arXiv:1509.01680 [cond-mat.str-el]].
\bibitem{GW86}
  D.~Gepner and E.~Witten,
  ``String Theory on Group Manifolds,''
  Nucl. Phys. B \textbf{278}, 493-549 (1986)
  doi:10.1016/0550-3213(86)90051-9
\bibitem{FO15}
  S.~C.~Furuya and M.~Oshikawa,
  ``Symmetry Protection of Critical Phases and a Global Anomaly in $1+1$ Dimensions,''
  Phys. Rev. Lett. \textbf{118}, no.2, 021601 (2017)
  doi:10.1103/PhysRevLett.118.021601
  [arXiv:1503.07292 [cond-mat.stat-mech]].
\bibitem{NY17}
  T.~Numasawa and S.~Yamaguchi,
  ``Mixed global anomalies and boundary conformal field theories,''
  JHEP \textbf{11}, 202 (2018)
  doi:10.1007/JHEP11(2018)202
  [arXiv:1712.09361 [hep-th]].
\bibitem{TS18}
  Y.~Tanizaki and T.~Sulejmanpasic,
  ``Anomaly and global inconsistency matching: $\theta$-angles, $SU(3)/U(1)^2$ nonlinear sigma model, $SU(3)$ chains and its generalizations,''
  Phys. Rev. B \textbf{98}, no.11, 115126 (2018)
  doi:10.1103/PhysRevB.98.115126
  [arXiv:1805.11423 [cond-mat.str-el]].
\bibitem{YHO18}
  Y.~Yao, C.~T.~Hsieh and M.~Oshikawa,
  ``Anomaly matching and symmetry-protected critical phases in $SU(N)$ spin systems in 1+1 dimensions,''
  Phys. Rev. Lett. \textbf{123}, no.18, 180201 (2019)
  doi:10.1103/PhysRevLett.123.180201
  [arXiv:1805.06885 [cond-mat.str-el]].
\bibitem{KY19}
  K.~Kikuchi and Y.~Zhou,
  ``Two-dimensional Anomaly, Orbifolding, and Boundary States,''
  [arXiv:1908.02918 [hep-th]].
\bibitem{LHYO22}
  L.~Li, C.~T.~Hsieh, Y.~Yao and M.~Oshikawa,
  ``Boundary conditions and anomalies of conformal field theories in 1+1 dimensions,''
  Phys. Rev. B \textbf{110}, no.4, 045118 (2024)
  doi:10.1103/PhysRevB.110.045118
  [arXiv:2205.11190 [hep-th]].
\bibitem{GKO86}
  P.~Goddard, A.~Kent and D.~I.~Olive,
  ``Unitary Representations of the Virasoro and Supervirasoro Algebras,''
  Commun. Math. Phys. \textbf{103}, 105-119 (1986)
  doi:10.1007/BF01464283
\bibitem{KW88}
  V.~G.~Kac and M.~Wakimoto,
  ``Modular and conformal invariance constraints in representation theory of affine algebras,''
  Adv. Math. \textbf{70}, 156 (1988)
  doi:10.1016/0001-8708(88)90055-2
\bibitem{K03}
  S.~K\"oster, ``Local Nature of Coset Models,''
  Reviews in Mathematical Physics 16 (2004) 353-382
  doi:10.1142/S0129055X0400200X
  [arXiv:math-ph/0303054].
\bibitem{R92}
  F.~Ravanini,
  ``Thermodynamic Bethe ansatz for G(k) x G(l) / G(k+l) coset models perturbed by their phi(1,1,Adj) operator,''
  Phys. Lett. B \textbf{282}, 73-79 (1992)
  doi:10.1016/0370-2693(92)90481-I
  [arXiv:hep-th/9202020 [hep-th]].
\bibitem{P88}
  V.~B.~Petkova,
  ``Two-dimensional (Half) Integer Spin Conformal Theories With Central Charge $C < 1$,''
  Int. J. Mod. Phys. A \textbf{3}, 2945-2958 (1988)
  doi:10.1142/S0217751X88001235
\bibitem{HNT20}
  C.~T.~Hsieh, Y.~Nakayama and Y.~Tachikawa,
  ``Fermionic minimal models,''
  Phys. Rev. Lett. \textbf{126}, no.19, 195701 (2021)
  doi:10.1103/PhysRevLett.126.195701
  [arXiv:2002.12283 [cond-mat.str-el]].
\bibitem{K20}
  J.~Kulp,
  ``Two More Fermionic Minimal Models,''
  JHEP \textbf{03}, 124 (2021)
  doi:10.1007/JHEP03(2021)124
  [arXiv:2003.04278 [hep-th]].
\bibitem{Cardy17}
  J.~Cardy,
  ``Bulk Renormalization Group Flows and Boundary States in Conformal Field Theories,''
  SciPost Phys. \textbf{3}, no.2, 011 (2017)
  doi:10.21468/SciPostPhys.3.2.011
  [arXiv:1706.01568 [hep-th]].
\bibitem{EGNO15}
  P.~Etingof, S.~Gelaki, D.~Nikshych and V.~Ostrik, ``Tensor Categories,'' American Mathematical Society, 2015.
\bibitem{M01}
  M.~Mueger, ``From Subfactors to Categories and Topology II. The quantum double of tensor categories and subfactors,'' Journal of Pure and Applied Algebra 180 (2003), 159-219 https://doi.org/10.1016/S0022-4049(02)00248-7
  [arXiv:math/0111205 [math.CT]].
\bibitem{DGNO09}
  V.~Drinfeld, S.~Gelaki, D.~Nikshych and V.~Ostrik,
  ``On braided fusion categories I,'' Selecta Mathematica 16 (2010), 1-119 https://doi.org/10.1007/s00029-010-0017-z
  [arXiv:0906.0620 [math.QA]].
\bibitem{T16}
  V.~G.~Turaev, ``Quantum Invariants of Knots and 3-Manifolds,''
  De Gruyter Studies in Mathematics, volume 18
  https://doi.org/10.1515/9783110435221.
\bibitem{P95}
  B.~Pareigis, ``On Braiding and Dyslexia,'' Journal of Algebra 171(1995), 413-425. https://doi.org/10.1006/jabr.1995.1019
\bibitem{BE99}
  J.~Bockenhauer and D.~E.~Evans,
  ``On alpha induction, chiral generators and modular invariants for subfactors,''
  Commun. Math. Phys. \textbf{208}, 429-487 (1999)
  doi:10.1007/s002200050765
  [arXiv:math/9904109 [math.OA]].
\bibitem{BEK99}
  J.~Bockenhauer, D.~E.~Evans and Y.~Kawahigashi,
  ``Chiral structure of modular invariants for subfactors,''
  Commun. Math. Phys. \textbf{210}, 733-784 (2000)
  doi:10.1007/s002200050798
  [arXiv:math/9907149 [math]].
\bibitem{BEK00}
  J.~Bockenhauer, D.~E.~Evans and Y.~Kawahigashi,
  ``Longo-Rehren subfactors arising from $\alpha$-induction,''
  [arXiv:math/0002154 [math.OA]]
\end{thebibliography}
\end{document}